\documentstyle[preprint,aps]{revtex}

\begin{document}

%=======================================================================
%   TITLE PAGE
%=======================================================================

%-----------------------------------
%   Preprint Number
%-----------------------------------
\preprint{\tighten\vbox{
\hbox{\bf MADPH--96--973}
\hbox{\bf UR--1487}
\hbox{November 1996}
}}
 
%-----------------------------------
%   Title
%-----------------------------------
\title{
The Electric Dipole Moment of the Muon \\
in a Two Higgs Doublet Model}
 
%-----------------------------------
%   Authors
%-----------------------------------
\author{Vernon Barger$^a$, Ashok Das$^b$ and Chung Kao$^a$}

%-----------------------------------
%   Address
%-----------------------------------
\address{
$^a$Department of Physics, University of Wisconsin, 
Madison, WI 53706 \\
$^b$Department of Physics and Astronomy, University of Rochester, 
Rochester, NY 14627}

\maketitle

\thispagestyle{empty}
 
\bigskip

%-----------------------------------
%   Abstract
%-----------------------------------
\begin{abstract}

The electric dipole moment of the muon ($d_\mu$) 
is evaluated in a two Higgs doublet model 
with a softly broken discrete symmetry. 
The leading contributions from both one loop and two loop diagrams 
are considered.
For $\tan\beta \equiv |v_2|/|v_1|$ close to one, 
contributions from two loop diagrams involving the $t$ quark 
and the $W$ boson dominate;
while for $\tan\beta \agt 10$, 
contributions from two loop diagrams involving the $b$ quark 
and the $\tau$ lepton are dominant.
In these two regions, $d_\mu \approx (m_\mu/m_e) d_e$.
For $8 \agt \tan\beta \agt 4$, significant cancellation occurs 
among the contributions from two loop diagrams 
and the one loop contribution dominates for $\tan\beta \sim 7$.
For $\tan\beta \agt 15$, the calculated $d_\mu$ 
can be close to the reach of a recently proposed experiment 
at the Brookhaven National Laboratory. 

\end{abstract}

\pacs{PACS numbers: 13.40.Fn, 11.30.Er, 12.15.Cc, 14.80.Dq}
%

%=======================================================================
%   BEGIN MAIN TEXT
%=======================================================================

%-----------------------------------------------------------------------
%   I. Introduction
%-----------------------------------------------------------------------
\section{Introduction}

A non-self-conjugate particle with a non-zero spin 
can have a permanent electric dipole moment (EDM) 
if the combined transformation of charge conjugation (C) and parity (P) 
is not an exact symmetry. 
In the Standard Model (SM) of electroweak interactions, 
CP violation is generated by the Kobayashi-Maskawa (KM) phase. 
The electron EDM generated from the KM mechanism is about 
$8 \times 10^{-41} {\rm e} \cdot {\rm cm}$ \cite{SM1,SM2,SM3}. 
Similarly, the expected muon EDM in the SM is about 
$2\times 10^{-38} {\rm e}\cdot {\rm cm}$. 
These values are more than 13 orders of magnitude below 
the current experimental limits. 
Therefore, precise measurements of the electron and the muon EDMs 
might reveal new sources of CP violation beyond the SM. 

Several experiments have been carried out 
to search for a neutron EDM ($d_n$) and an electron EDM ($d_e$). 
At present, the experimental upper limits on $d_n$ \cite{dn} 
and $d_e$ \cite{de} at 95\% C.L. are
$|d_n| <  11 \times 10^{-26} {\rm e}\cdot {\rm cm}$ 
and  
$|d_e| <  6.2 \times 10^{-27} {\rm e}\cdot {\rm cm}$ 
respectively.
The measurement of the muon EDM ($d_\mu$) is not yet so precise 
as that for $d_e$ and $d_n$. 
The experimental upper limit on the muon EDM at 95\% C.L. is 
$|d_\mu| < 1.1 \times 10^{-18} {\rm e}\cdot {\rm cm}$ \cite{dmu}.
Recently, a dedicated experiment has been proposed \cite{Yannis} 
to measure the electric dipole moment of the muon with the existing $g-2$ ring 
at the Brookhaven National Laboratory (BNL) \cite{E821}.
This experiment may be able to improve the measurement of the muon EDM
by at least 4 orders of magnitude.

A significant electric dipole moment for the electron or the neutron 
can be generated if CP violation is mediated by Higgs-boson exchange 
\cite{Steve89,Steve90,Barr&Zee}. 
In a non-supersymmetric model with multi-Higgs doublets, 
for flavor symmetry to be conserved naturally to a good degree, 
a discrete symmetry is usually required \cite{Glashow&Weinberg}. 
In a two Higgs doublet model, 
there will be no CP violation from the Higgs sector 
if the discrete symmetry enforcing natural flavor conservation is exact.
If this symmetry is broken only by soft terms,
CP violation can be introduced and the flavor changing interaction 
can be kept at an acceptably low level as well \cite{Branco,Jiang}.

The EDM of an elementary fermion, generated from Higgs-boson exchange, 
has dominant contributions from one-loop or two-loop diagrams. 
Intuitively, the one-loop contributions 
are proportional to the product of two Yukawa couplings $(m/v)^2$;  
while the leading two-loop contributions are proportional to 
only one Yukawa coupling $m/v$, where $m$ is the fermion mass 
and $v = 246$ GeV is the vacuum expectation value (VEV) of the SM Higgs field. 
The exact one loop result actually goes like $m^3/m_0^2 \ln(m^2/m_0^2)$, 
where $m_0$ is the mass of the lightest neutral Higgs boson.
The two-loop contributions dominate for the electron EDM 
because its one-loop contributions are greatly suppressed.
For the muon EDM, the one-loop contributions could be comparable to 
those from two-loop diagrams for large $\tan\beta$.

Several studies have confirmed that a significant electron EDM 
can be generated if Higgs-boson exchange mediates CP violation 
in a two Higgs doublet model 
with Yukawa interactions of Model II \cite{Model2a,Model2b}.  
In this Model, 
one Higgs doublet couples to down-type quarks and charged leptons 
while another doublet couples to up-type quarks and neutrinos. 
The electron EDM has important contributions from two-loop diagrams 
involving the top quark \cite{Barr&Zee} and the $W$ boson 
\cite{Ruiming,Wpm1,Wpm2}. 
Diagrams with the charged Higgs boson ($H^\pm$) 
can also contribute to the electron EDM \cite{Kao&Xu}. 
However, recent measurements on the decay rate of $b \to s\gamma$ 
by the CLEO collaboration \cite{CLEO} 
constrain the mass of the $H^\pm$ in the Model II 
to be at least three times larger than the $W$ boson mass 
\cite{bsg1,bsg2,bsg3,bsg4,bsg5,bsg6}, 
and contributions from the $H^\pm$ are therefore not expected 
to be significant. 
For the muon EDM, the two loop contributions involving the $t$ quark 
as well as the $W$ boson were found to be dominant 
and to satisfy a simple scaling relation 
$d_\mu/d_e \simeq m_\mu/m_e$ \cite{Bernreuther,Suzuki}. 
However, this conclusion did not take into account potentially important 
two loop contributions involving the $b$-quark and the $\tau$-lepton.

In this paper, 
we present the first complete calculation for the muon EDM 
in a two Higgs doublet model with CP violation 
generated from Higgs-boson exchange and Yukawa couplings of Model II.
We evaluate the leading contributions to the EDM for the muon 
from two loop diagrams involving the $W$ boson, 
the $t$ quark, the $b$ quark and the $\tau$ lepton 
and the one loop diagrams.
We find that 
(i) for $\tan\beta \agt 10$ contributions 
from two loop diagrams involving the $b$ quark and the $\tau$ lepton 
are dominant \cite{Das&Kao}; 
(ii) for $8 \agt \tan\beta \agt 4$, 
significant cancellation occurs among the two loop diagrams 
and the one loop contribution dominates for $\tan\beta \sim 7$.

%-----------------------------------------------------------------------
%   II. CP Violation from Higgs Exchange
%-----------------------------------------------------------------------
\section{CP Violation from Higgs Exchange}

In a two Higgs doublet model with a discrete symmetry softly broken, 
the Higgs potential can have the following form \cite{Steve90} 
\begin{eqnarray}
V[\phi_1,\phi_2] 
& = &  m_1 \phi_1^{\dagger}\phi_1  +m_2 \phi_2^{\dagger}\phi_2
     +\eta \phi_1^{\dagger}\phi_2+\eta^*\phi_2^{\dagger}\phi_1
      \nonumber \\
&   &+\frac{1}{2} g_1 (\phi_1^{\dagger}\phi_1)^2 
     +\frac{1}{2} g_2 (\phi_2^{\dagger}\phi_2)^2
     \nonumber \\
&   &+g  (\phi_1^{\dagger}\phi_1)(\phi_2^{\dagger}\phi_2)
     +g' (\phi_1^{\dagger}\phi_2)(\phi_2^{\dagger}\phi_1)
     +\frac{1}{2} h   (\phi_1^{\dagger}\phi_2)^2
     +\frac{1}{2} h^* (\phi_2^{\dagger}\phi_1)^2. 
\label{eq:Higgs1}
\end{eqnarray}
This potential respects a discrete symmetry 
$\phi_1 \to -\phi_1$, and $\phi_2 \to +\phi_2$, 
except for the soft terms $\phi_1^\dagger \phi_2$ and $\phi_1^\dagger \phi_2$, 
We can rewrite the Higgs potential as 
\begin{eqnarray}
V[\phi_1,\phi_2] 
& = &+\frac{1}{2} g_1 (\phi_1^{\dagger}\phi_1 -\frac{|v_1|^2}{2})^2 
     +\frac{1}{2} g_2 (\phi_2^{\dagger}\phi_2 -\frac{|v_2|^2}{2})^2 
     \nonumber \\
&   &+g  (\phi_1^{ \dagger}\phi_1 -\frac{|v_1|^2}{2} )
         (\phi_2^{ \dagger}\phi_2 -\frac{|v_2|^2}{2} )
     +g' |\phi_1^{\dagger}\phi_2 -\frac{v_1^*v_2}{2} |^2 \nonumber \\
&   &+\frac{1}{2} h  (\phi_1^{\dagger}\phi_2 -\frac{v_1^*v_2}{2})^2
     +\frac{1}{2} h^*(\phi_2^{\dagger}\phi_1 -\frac{v_2^*v_1}{2})^2 
     \nonumber \\
&   &+\xi (\frac{\phi_1^{\dagger}}{v_1^*} -\frac{\phi_2^{\dagger}}{v_2^*}) 
          (\frac{\phi_1}{v_1} -\frac{\phi_2}{v_2}), 
\label{eq:Higgs2}
\end{eqnarray}
where all the coupling constants are real, with the possible exception of $h$. 
CP is violated if $h v_1^{*2} v_2^2$ has an imaginary part.
The minimum of this potential occurs at  
\begin{equation}
<\phi_1> = {\frac {v_1}{\sqrt{2}}}, \;\; 
<\phi_2> = {\frac {v_2}{\sqrt{2}}}.
\end{equation}
where $v_1/\sqrt{2}$ and $v_2/\sqrt{2}$ 
are the vacuum expectation values (VEVs) of $\phi_1$ and $\phi_2$. 
Both VEVs can be complex. 
Without loss of generality, we will take 
$<\phi^0_1> = v_1/\sqrt{2}$ and $<\phi^0_2> = v_2 e^{i\theta}/\sqrt{2}$ 
with $v_1$ and $v_2$ real and $\tan\beta \equiv v_2/v_1$.  
The two VEVs satisfy $\sqrt{v_1^2 +v_2^2} = v$, where $v$ is 
the VEV of the SM Higgs field and $v^2 = (\sqrt{2} G_F)^{-1}$.

Introducing a transformation, which takes the two Higgs doublets 
to the eigenstates $\Phi_1$ and $\Phi_2$, 
such that $<\Phi^0_1> = v/\sqrt{2}$ and $<\Phi^0_2> = 0$, we have 
\begin{eqnarray}
\phi_1 &=& \cos \beta \Phi_1 -\sin \beta \Phi_2, \,\,
\phi_2 = ( \sin \beta \Phi_1 +\cos \beta \Phi_2 ) e^{i\theta} , \nonumber \\
\Phi_1  &=& 
\left( \begin{array}{c} 
G^+ \\ \frac{v +H_1 +iG^0}{\sqrt{2}} 
\end{array} \right), \,\,
\Phi_2   =  
\left( \begin{array}{c} 
H^+ \\ \frac{ H_2 +iA}{\sqrt{2}} 
\end{array} \right), 
\label{eq:gauge}
\end{eqnarray}
where $G^\pm$ and $G^0$ are Goldstone bosons,
$H^\pm$ are singly charged Higgs bosons,
$H_1$ and $H_2$ are CP-even scalars, 
and $A$ is a CP-odd pseudoscalar.

In the new eigenstates of Eq.~\ref{eq:gauge} for the Higgs fields, 
the Higgs potential becomes \cite{Kao&Xu}, 
\begin{eqnarray}
V [\Phi_1,\Phi_2] 
& = & \frac{1}{2}\lambda_1(\Phi_1^{\dagger}\Phi_1-{\frac {v^2}{2}})^2 
     +\frac{1}{2}\lambda_2(\Phi_2^{\dagger}\Phi_2)^2
     \nonumber \\
&   &+\lambda_3(\Phi_1^{\dagger}\Phi_1-{\frac {v^2}{2}})\Phi_2^{\dagger}\Phi_2 
     +\lambda_4(\Phi_1^{\dagger}\Phi_2)(\Phi_2^{\dagger}\Phi_1)
     \nonumber \\
&   &+\lambda_5(\Phi_1^{\dagger}\Phi_1+\Phi_2^{\dagger}\Phi_2-{\frac {v^2}{2}})
      (\Phi_1^{\dagger}\Phi_2+\Phi_2^{\dagger}\Phi_1) 
     \nonumber \\
&   &+(\lambda_6 \Phi_1^{\dagger}\Phi_2 +\lambda_6^* \Phi_2^{\dagger}\Phi_1)
      (\Phi_1^{\dagger}\Phi_1-\Phi_2^{\dagger}\Phi_2-{\frac {v^2}{2}}) 
     \nonumber \\
&   &+\frac{1}{2}\lambda_7  (\Phi_1^{\dagger}\Phi_2)^2
     +\frac{1}{2}\lambda_7^*(\Phi_2^{\dagger}\Phi_1)^2 \nonumber\\
&   &+\rho (\Phi_2^{\dagger}\Phi_2), 
\label{eq:Higgs3}
\end{eqnarray}
where the parameters $\rho$, $v$ and $\lambda_i$, $i=1$ through $5$, 
are all real; $\lambda_6$ and $\lambda_7$ can be complex.
CP is violated if the imaginary part of $\lambda_6$ or $\lambda_7$ 
is nonvanishing.  
There are 11 parameters, but only 10 of them are independent. 
One of the relations among the 11 parameters is
\begin{equation}
{\rm Im}(\lambda_6) = 
-{\frac {\lambda_5}{2 (\lambda_1-\lambda_2)}} {\rm Im}(\lambda_7).
\end{equation}
In the above parameterization, $\tan \beta$ is given by 
\begin{equation}
\tan\beta = {\frac {\lambda_1-\lambda_2}{\lambda_5}}
+\sqrt{1+{\frac {(\lambda_1-\lambda_2)^2}{\lambda_5^2}}}.
\end{equation}

In Model II \cite{Model2a,Model2b}, 
the down-type quarks and charged leptons have bilinear couplings 
with the doublet $\phi_1$, 
while all the up-type quarks and neutrinos have bilinear couplings 
with the doublet $\phi_2$.
The Lagrangian density of Yukawa interactions has the following form 
\begin{eqnarray}
{\cal L}_Y 
& = & -\sum_{m,n=1}^{3} \bar{L}^m_L \phi_1 E_{mn} l^n_R
      -\sum_{m,n=1}^{3} \bar{Q}^m_L \phi_1 F_{mn} d^n_R
       \nonumber \\
&   & -\sum_{m,n=1}^{3} \bar{Q}^m_L \tilde{\phi_2} G_{mn} u^n_R 
       \,\, +{\rm H.c.}, 
\end{eqnarray}
where
\begin{equation}
\phi_\alpha 
= \left( \begin{array}{c} \phi_\alpha^+ \\ 
                          \frac{ v_\alpha +\phi_\alpha^0 }{ \sqrt{2} }
         \end{array} \right), \,\, 
\tilde{\phi}_\alpha 
= \left( \begin{array}{c} \frac{ v_\alpha^* +{\phi_\alpha^0}^* }{ \sqrt{2} } \\ 
                         -\phi_\alpha^- 
         \end{array} \right), \,\, 
\phi_\alpha^- = {\phi_\alpha^+}^*, \,\, \alpha = 1,2, 
\end{equation}
and
\begin{equation}
L^m_L = \left( \begin{array}{c} \nu_l \\ l \end{array} \right)^m_L, \,\, 
Q^m_L = \left( \begin{array}{c} u \\ d \end{array} \right)^m_L, \,\, 
m= 1,2,3, 
\end{equation}
$l^m$, $d^m$, and $u^m$ are the leptons, the down-type quarks 
and the up-type quarks in the gauge eigenstates.
This Lagrangian respects a discrete symmetry,
\begin{eqnarray}
\phi_1 &\to& -\phi_1, \,\, \phi_2 \to +\phi_2, \nonumber \\
l^m_R  &\to& -l^m_R,  \,\, d^m_R  \to -d^m_R,  \nonumber \\
L^m_L  &\to& +L^m_L,  \,\, Q^m_L  \to +Q^m_L, \,\, u^m_R  \to +u^m_R, 
\end{eqnarray}
with $m = 1,2,3$ and $\alpha = 1,2$.

In the new eigenstates of Eq.~\ref{eq:gauge} for the Higgs fields, 
the neutral Yukawa interactions of the quarks become 
\begin{eqnarray}
{\cal L}^N_Y
& = & -\sum_{l=e,\mu,\tau}\frac{m_l}{v} \bar{l}l ( H_1-\tan\beta H_2 ) 
     -i\sum_{l=e,\mu,\tau}\frac{m_l}{v} \bar{l}\gamma_5 l ( G^0-\tan\beta A ) 
       \nonumber \\
&   & -\sum_{d=d,s,b}\frac{m_d}{v} \bar{d}d ( H_1-\tan\beta H_2 ) 
     -i\sum_{d=d,s,b}\frac{m_d}{v} \bar{d}\gamma_5 d ( G^0-\tan\beta A ) 
       \nonumber \\
&   & -\sum_{u=u,c,t} \frac{m_u}{v}\bar{u}u [ H_1+\cot\beta H_2 ] 
     +i\frac{m_u}{v}\bar{u} \gamma_5 u [ G^0+\cot\beta A ].
\end{eqnarray}
where the quarks and leptons are in the mass eigenstates.

Adopting Weinberg's parameterization \cite{Steve90} 
we can write the following neutral Higgs-boson exchange propagators as 
\begin{eqnarray}
< H_1 A >_q 
& = & \frac{1}{2} \sum_n \frac{ \sin 2\beta {\rm Im}Z_{0n} }{ q^2-m_n^2 } 
  =   \frac{1}{2} \sum_n 
      \frac{-\cos^2 \beta \cot \beta{\rm Im}\tilde{Z}_{1n}
            +\sin^2 \beta \tan \beta{\rm Im}\tilde{Z}_{2n} }{ q^2-m_n^2 },
    \nonumber \\
< H_2 A >_q
& = & \frac{1}{2} \sum_n 
      \frac{ \cos 2\beta {\rm Im}Z_{0n} -{\rm Im}\tilde{Z}_{0n} }{ q^2-m_n^2 }
  =   \frac{1}{2} \sum_n 
      \frac{ \cos^2 \beta {\rm Im}\tilde{Z}_{1n} 
            +\sin^2 \beta {\rm Im}\tilde{Z}_{2n} }{ q^2-m_n^2 },
\label{eq:H&A}
\end{eqnarray}
where the summation is over all the mass eigenstates of neutral Higgs 
bosons. We approximate the above expressions by assuming that 
the sums are dominated by the lightest neutral Higgs boson of mass $m_0$, 
and drop the sums and indices $n$ in Eq.~\ref{eq:H&A} hereafter.
There are relations among the CP violation parameters: 
\begin{eqnarray}
{\rm Im} Z_0 +{\rm Im} \tilde{Z}_0 
& = & -\cot^2 \beta {\rm Im} \tilde{Z}_1, \nonumber \\
{\rm Im} Z_0 -{\rm Im} \tilde{Z}_0 
& = & +\tan^2 \beta {\rm Im} \tilde{Z}_2. 
\label{eq:ImZ1}
\end{eqnarray}
Employing unitarity, Weinberg has shown \cite{Steve90} that 
$|{\rm Im} \tilde{Z}_1| \le (1/2) \tan\beta ( 1+\tan^2\beta)^{1/2}$ 
and 
$|{\rm Im} \tilde{Z}_2| \le (1/2) \cot\beta ( 1+\cot^2\beta)^{1/2}$. 
Therefore, we have 
\begin{eqnarray}
|{\rm Im}Z_0 +{\rm Im}\tilde{Z}_0| 
& \le & (1/2) \cot\beta ( 1+\tan^2\beta)^{1/2}, \nonumber \\ 
|{\rm Im}Z_0 -{\rm Im}\tilde{Z}_0| 
& \le & (1/2) \tan\beta ( 1+\cot^2\beta)^{1/2}. 
\label{eq:Unitarity}
\end{eqnarray}
%

%-----------------------------------------------------------------------
%   III. Muon EDM
%-----------------------------------------------------------------------
\section{Muon Electric Diople Moment}

%-----------------------------------------------------------------------
%	 The Two Loop Diagrams
%-----------------------------------------------------------------------
\subsection{Two Loop Diagrams}

Two loop diagrams for fermion loops contributing to the muon EDM 
are illustrated in Figures 1(c) and 1(d).
The diagrams with the intermediate $Z$ boson are highly suppressed 
by the vector part of the $Z\mu^+\mu^-$ couplings.  
Therefore, we consider only the diagrams 
involving an intermediate $\gamma$.

The muon EDM generated from two loop diagrams 
with the top quark is \cite{Barr&Zee} 
\begin{equation}
d_\mu^{ t-{\rm loop} }
=  -\frac{16}{3}
    \frac{ e m \alpha \sqrt{2} G_F }{ (4\pi)^3 }
    \{ [ f(\rho_t) +g(\rho_t) ] {\rm Im}Z_0 
      +[ g(\rho_t) -f(\rho_t) ] {\rm Im}\tilde{Z}_0 \}, 
\label{eq:top}
\end{equation}
where $m = m_\mu$, $\rho_t = m_t^2/m_0^2$, 
and $m_0$ is {\it the mass of the lightest neutral Higgs boson}. 
The functions $f$ and $g$ are defined as
\begin{eqnarray}
f(r) & \equiv & \frac{r}{2} \int_0^1 dx \frac{ 1-2x(1-x) }{ x(1-x)-r }
                \ln\left[ \frac{ x(1-x) }{r} \right], \nonumber \\
g(r) & \equiv & \frac{r}{2} \int_0^1 dx \frac{1}{ x(1-x)-r } 
                \ln\left[ \frac{ x(1-x) }{r} \right]. 
\label{eq:f&g}
\end{eqnarray}
We take $m_t = 175$ GeV, $m_b = 4.8$ GeV, 
$m_\tau = 1.777$ GeV, $m_W = 80.0$ GeV, 
and the fine structure constant $\alpha = 1/137$, 
neglecting the running up to the scale of $m_0$.

The EDM generated from the $b$ and the $\tau$ loops is \cite{Das&Kao}
\begin{equation}
d_\mu^{ b,\tau-{\rm loop} } 
= -( 4N_c Q^2 \tan^2\beta )
   \frac{ e m \alpha \sqrt{2} G_F }{ (4\pi)^3 }
   [ f(\rho_f) +g(\rho_f) ] ( {\rm Im}Z_0 +{\rm Im}\tilde{Z}_0 ),
\label{eq:bottom}
\end{equation}
where $N_c$ is the color factor and $Q$ is the charge.
For the $b$ and the $\tau$, $4N_c Q^2$ is equal to
$4/3$ and $4$ respectively. 

The leading contribution from two loop diagrams 
with the $W$ boson \cite{Ruiming,Wpm1,Wpm2} is
\begin{eqnarray}
d_1^{ W-{\rm loop} } 
& = & ( \sin^2\beta )
      \frac{ e m \alpha \sqrt{2} G_F }{ (4\pi)^3 } 
      [ 4 I_1(\rho_W) +2 I_2(\rho_W) ] {\rm Im} Z_0, \\
I_1(\rho_W) & = & 3 f(\rho_W)+\frac{23}{4} g(\rho_W)+\frac{3}{4} h(\rho_W), \\
I_2(\rho_W) & = & \frac{ f(\rho_W)-g(\rho_W) }{\rho_W}, \\
h(r) & \equiv & 
     \frac{r}{2} \int_0^1 dx \frac{1}{ x(1-x)-r } 
     \left[ \frac{r}{x(1-x)-r}\ln\left( \frac{ x(1-x) }{r} \right) -1 \right], 
\label{eq:wpm}
\end{eqnarray}
where $\rho_W = m_W^2/m_0^2$, 
and the functions $f$ and $g$ are defined in Eq.~\ref{eq:f&g}.
There are another two sets of diagrams with the $W$ boson \cite{Ruiming} 
contributing to the muon EDM. 
The contribution from these additional diagrams 
has an opposite sign to that of $d_1^{ W-{\rm loop} }$ 
and therefore reduces the magnitude of the full contribution 
from the $W$ boson.
In our analysis, we have employed the formulas in Ref.~\cite{Ruiming} 
to evaluate the complete contributions from the $W$ loops.

%-----------------------------------------------------------------------
%	 The One Loop Diagrams
%-----------------------------------------------------------------------
\subsection{One Loop Diagrams}

The one loop diagrams contributing to the muon EDM 
are illustrated in Figures 1(a) and 1(b).
The one-loop contribution to the muon EDM ($d_\mu$) is 
\begin{equation}
d_\mu^{1-{\rm loop}}
 =  \frac{ e m \sqrt{2} G_F \tan^2\beta }{ (4\pi)^2 }
    I(\rho)( {\rm Im}Z_0 +{\rm Im}\tilde{Z}_0 ),
\label{eq:Muon1}
\end{equation}
where $\rho = m^2/m_0^2$.
The function $I(\rho)$ is defined as  
\begin{eqnarray}
I(\rho)
& \equiv & \rho \int_0^1 dx \frac{x^2}{ \rho x^2 -x +1 } \nonumber \\
& = & 1 +\frac{1}{ \rho (z_1-z_2) }
        [ (z_1-1)\ln(\frac{z_1-1}{z_1})
         -(z_2-1)\ln(\frac{z_2-1}{z_2}) ], 
\label{eq:Irho}
\end{eqnarray}
where $z_1$ and $z_2$ are roots of $\rho x^2 -x +1 = 0$. 
For $\rho << 1$, $I(\rho)$ approaches $ -\rho [ \ln(\rho) +3/2 ]$.
In this limit, the one-loop contribution 
to the the muon or the electron EDM has a simple form 
\begin{equation}
d_{(e,\mu)}^{1-{\rm loop}}
 = -\frac{ e \sqrt{2} G_F \tan^2\beta }{ (4\pi)^2 } 
    (m^3/m_0^2) [ \ln(m^2/m_0^2) +3/2 ]
    ( {\rm Im}Z_0 +{\rm Im}\tilde{Z}_0 ).
\label{eq:Muon1b}
\end{equation}
where $m = m_e, m_\mu$ and ($m_0 >> m$) is assumed. 
Therefore, the one-loop contribution to the muon 
or the electron EDM is proportional to $(m^3/m_0^2) \ln(m^2/m_0^2)$.

%-----------------------------------------------------------------------
%	 The Muon EDM
%-----------------------------------------------------------------------
\subsection{Numerical Values}

In this section, we discuss the numerical value of the muon EDM 
with several choices of $m_0$ and $\tan\beta$.
The contributions from one loop diagrams (1-loop) 
and two loop diagrams involving 
the $W$ boson ($W$-loop), the $t$ quark ($t$-loop), 
the $b$ quark ($b$-loop) and the $\tau$ lepton ($\tau$-loop), 
are presented in Table I for $\tan\beta =$ 2 and 20
with $m_0 = 100$, 200 and 400 GeV, 
in units of (a) Im $Z_0$ and (b) Im $\tilde{Z}_0$. 

To study the muon EDM dependence on the lightest Higgs boson mass ($m_0$), 
we present the muon EDM generated from one loop and two loop diagrams 
in units of (a) Im $Z_0$ and (b) Im $\tilde{Z}_0$, as a function of $m_0$, 
with $\tan\beta =$ 1, 7 and 20 in Figs. 2, 3 and 4. 
Figure 5 shows the effect of varying $\tan\beta$ on the muon EDM, 
for the case of $m_0 = 100$ GeV.
From these Figures, we conclude that 
(i) for $\tan\beta$ close to one, 
contributions from two loop diagrams involving the top quark 
and the $W$ boson dominate; 
(ii) while for $\tan\beta \agt 10$ contributions 
from two loop diagrams involving the $b$ quark and the $\tau$ lepton 
are dominant \cite{Das&Kao}; 
(iii) for $8 \agt \tan\beta \agt 4$, significant cancellation occurs 
among the contributions from two loop diagrams 
and one loop diagrams dominate for $\tan \sim 7$.

To compare the muon EDM with the electron EDM, 
we present the ratio of ($d_\mu/d_e$) to ($m_\mu/m_e$) in Fig. 6.
For the electron EDM, the two-loop contribution 
is about $10^5({\rm for} \tan\beta \agt 10)$  
to $10^6({\rm for} \tan\beta \sim 1)$ 
larger than that from the one-loop diagrams.
For the muon EDM, the one loop diagrams 
make important contributions which dominate for $\tan\beta \sim 7$.
For $\tan\beta \agt 10$, the one-loop contribution to the muon EDM 
is about $10\%$ of that from the two-loop diagrams.
We find that 
(i) for $\tan\beta \sim 1$,  $d_\mu \approx (m_\mu/m_e)$; 
(ii) for $\tan\beta \agt 10$, $d_\mu \approx 0.9 (m_\mu/m_e)$, 
since there is a cancellation for the muon EDM 
between the one-loop contribution and 
the two-loop contribution involving the $b$-quark and the $\tau$-lepton; 
(iii) for $8 \agt \tan\beta \agt 4$, 
$|d_\mu|$ can be two to three times $|(m_\mu/m_e) d_e|$.

The muon EDM can be expressed as
\begin{eqnarray}
d_\mu & = & d_A {\rm Im} Z_0 +d_B {\rm Im} \tilde{Z}_0 \nonumber \\
      & = & \frac{d_A+d_B}{2} ({\rm Im} Z_0 +{\rm Im} \tilde{Z}_0 ) 
           +\frac{d_A-d_B}{2} ({\rm Im} Z_0 -{\rm Im} \tilde{Z}_0 )
\end{eqnarray}
where $d_A$ and $d_B$ are the total coefficients of ${\rm Im} Z_0$ 
and ${\rm Im} \tilde{Z}_0$ from one loop diagrams, 
the $W$-loop, the $t$-loop, the $b$-loop, and the $\tau$-loop. 
Applying unitarity constraints in Eq.~\ref{eq:Unitarity}, 
we can define the maximal $|d_\mu|$ as 
\begin{eqnarray}
|d_\mu|_{\rm MAX}
& = & |\frac{d_A+d_B}{2}| |{\rm Im} Z_0 +{\rm Im} \tilde{Z}_0| 
     +|\frac{d_A-d_B}{2}| |{\rm Im} Z_0 -{\rm Im} \tilde{Z}_0| 
      \nonumber \\
& = & \frac{|d_A+d_B|}{4} \cot\beta ( 1+\tan^2\beta )^{1/2}
     +\frac{|d_A-d_B|}{4} \tan\beta ( 1+\cot^2\beta )^{1/2}. 
\label{eq:dMAX}
\end{eqnarray}

In Figure 7, we present the maximal value allowed by unitarity 
for the muon and the electron EDMs, 
as a function of $\tan\beta$ with $m_0 = 100$, 200 and 400 GeV. 
Also shown is the experimental upper limit 
for the electron EDM (denoted by $|d_e|_{\rm U.L.}$).
In this model, for $\tan\beta \agt 10$, 
the simple scaling $(m_\mu/m_e)|d_e|_{\rm U.L.}$ could be treated 
as an upper limit for the muon EDM since $d_\mu \approx (m_\mu/m_e) d_e$. 
A measured value of the muon EDM above this bound could arise
if the muon EDM and the electron EDM are generated from different sources,
$i.$ $e.$, if this model applies only to the muon EDM.

There are several interesting aspects to note 
from the different contributions: 
(i) The contributions from one loop diagrams and two loop diagrams 
with the $b$ and the $\tau$ are proportional to (Im $Z_0$ +Im $\tilde{Z}_0$).
(ii) For the the $t$-loop, the coefficient of the Im $\tilde{Z}_0$ 
is much smaller than that of the Im $Z_0$.
(iii) The $W$-loop does not contribute to the Im $\tilde{Z}_0$ term.
(iv) The contributions from one-loop diagrams and the $W$-loop 
have positive sign while the contributions from heavy fermion loops 
have negative sign.
Therefore, the total muon EDM is slightly reduced by cancellation.
(v) For large $\tan\beta$, 
two loop diagrams involving the $b$ and the $\tau$ dominate 
and the electron EDM and the muon EDM are almost proportional to their masses.
Therefore, $d_\mu \approx (m_\mu/m_e) d_e$ for $\tan\beta \agt 10$.

%-----------------------------------------------------------------------
%   IV. Conclusions
%-----------------------------------------------------------------------
\section{Conclusions}

Our results may be summarized as follows:
\begin{itemize}
\item For $\tan\beta$ close to one, 
contributions from two loop diagrams involving the $t$ quark 
and the $W$ boson dominate.
\item For $\tan\beta \agt 10$, 
contributions from two loop diagrams involving the $b$ quark 
and the $\tau$ lepton are dominant.
\item For $\tan\beta \sim 1$ or $\tan\beta \agt 10$, 
$d_\mu \approx (m_\mu/m_e) d_e$.
\item For $8 \agt \tan\beta \agt 4$, significant cancellations occur 
among the contributions from two loop diagrams 
and the one loop contribution dominates for $\tan\beta \sim 7$.
\item For $\tan\beta > 15$ and the lightest Higgs boson mass $m_0 < 300$ GeV, 
CP violation mediated by Higgs-boson exchange in a two Higgs doublet model 
could produce a muon EDM which is close to the reach of 
the proposed BNL experiment \cite{Yannis}.
\end{itemize}

High precision experimental measurements of the muon EDM 
could provide to interesting information about $m_0$ and $\tan\beta$ 
as well as CP violation parameters, 
${\rm Im}Z_0$, and ${\rm Im}\tilde{Z}_i, i = 0,1,2$. 
A positive result for the muon EDM measurement 
could shed light on non-standard CP violation 
and help pin down the value for $\tan\beta$ in two Higgs doublet models. 
A negative result, however, could mean 
(i) that $\tan\beta$ is small than 10; or,
(ii) the CP violation parameters ${\rm Im}Z_0$, and ${\rm Im}\tilde{Z}_i$ 
are smaller than their unitarity bounds; or, 
(iii) the masses of the neutral Higgs scalars and the Higgs pseudoscalar are 
very close to one another \cite{Steve90}.
 
%------------------------------------------------------------------------------
%       THE ACKNOWLEDGEMENTS
%------------------------------------------------------------------------------
\section*{Acknowledgments}

We are grateful to Yannis Semertzidis for continuing encouragement 
and beneficial discussions regarding the BNL muon EDM experiment.
We also thank Robert Garisto for a helpful comment.
This research was supported in part by the U.S. Department of Energy 
under Grants 
No. DE-FG05-87ER40319 (Rochester) and 
No. DE-FG02-95ER40896 (Wisconsin),
and in part by the University of Wisconsin Research Committee 
with funds granted by the Wisconsin Alumni Research Foundation.
 
%-----------------------------------------------------------------------
%   REFERENCES
%-----------------------------------------------------------------------
%

%-----------------------------------------------------------------------
%   TABLE CAPTIONS
%-----------------------------------------------------------------------
\newpage
\begin{table}
\caption[]{
The muon EDM from one-loop diagrams (1-loop) 
and two-loop diagrams with the $W$ boson ($W$-loop), 
the top quark ($t$-loop), the bottom quark ($b$-loop), 
the $\tau$-lepton ($\tau$-loop) and their total (Total) 
for $\tan\beta = 2$ ($\tan\beta = 20$)
with the lightest Higgs boson mass $m_0 = 100$, 200 and 400 GeV, 
in units of (a)~${\rm Im} Z_0 \times 10^{-24} {\rm e}\cdot {\rm cm}$ 
and (b) ${\rm Im} \tilde{Z}_0 \times 10^{-24} {\rm e}\cdot {\rm cm}$.}

\bigskip

\begin{tabular}{clll} 
         & \multicolumn{3}{c}{$m_0$ (GeV)} \\
\cline{2-4} 
Diagrams & \multicolumn{1}{c}{100} 
         & \multicolumn{1}{c}{200}  & \multicolumn{1}{c}{400} \\
\tableline
(a) Im $Z_0 \times 10^{-24} {\rm e}\cdot {\rm cm}$ \\
1-loop      & $+0.01$ ($+1.19$)& $+0.003$ ($+0.33$) & $+0.001$ ($+0.09$) \\
$W$-loop    & $+2.47$ ($+3.07$)& $+1.16$  ($+1.45$) & $+0.20$  ($+0.25$) \\
$t$-loop    & $-1.88$ ($-1.88$)& $-1.23$  ($-1.23$) & $-0.72$  ($-0.72$) \\
$b$-loop    & $-0.06$ ($-5.61$)& $-0.021$ ($-2.08$) & $-0.007$  ($-0.73$) \\
$\tau$-loop & $-0.04$ ($-3.97$)& $-0.014$ ($-1.36$) & $-0.004 $ ($-0.44$) \\
Total       & $+0.50$  ($-7.2$)& $-0.10$  ($-2.8$)  & $-0.53$  ($-1.5$)   \\
\tableline
(b) Im $\tilde{Z}_0 \times 10^{-24} {\rm e}\cdot {\rm cm}$ \\
1-loop      & $+0.01$ ($+1.19$)& $+0.003$ ($+0.33$) & $+0.001$ ($+0.09$)\\
$W$-loop    & $+0.0$           & $+0.0$             & $+0.0$ \\
$t$-loop    & $-0.33$ ($-0.33$)& $-0.21$  ($-0.21$) & $-0.11$  ($-0.11$)\\
$b$-loop    & $-0.06$ ($-5.61$)& $-0.021$ ($-2.08$) & $-0.007$  ($-0.73$) \\
$\tau$-loop & $-0.04$ ($-3.97$)& $-0.014$ ($-1.36$) & $-0.004 $ ($-0.44$) \\
Total       & $-0.42$ ($-8.7$) & $-0.24$  ($-3.3$)  & $-0.13$  ($-1.2$)
\end{tabular}
\end{table}
%

%-----------------------------------------------------------------------
%   FIGURE CAPTIONS
%-----------------------------------------------------------------------

% FIG. 1
\begin{figure}
\caption[]{
Feynman diagrams for leading contributions to the muon EDM 
from one loop diagrams [(a) and (b)]
and two loop diagrams with heavy fermions and the $W$ boson [(c) and (d)]. 
There are many more diagrams involving the $W$ boson 
\cite{Ruiming,Wpm1,Wpm2} 
that are not shown in this figure.
\label{fig:EDM}
}\end{figure}

% FIG. 2
\begin{figure}
\caption[]{
The muon EDM generated from one-loop diagrams (dash), 
two-loop diagrams (dash-dot) and their total (solid) 
in units of (a) Im $Z_0$ and (b) Im $\tilde{Z}_0$,  
as a function of $m_0$, with $\tan\beta = 1$, 
where $m_0$ is the mass of the lightest neutral Higgs boson. 
Also shown are the contributions from two-loop diagrams 
involving the $W$ boson (dot),
the $t$ quark (dash-dot),
the $b$ quark (dash-dot-dot), 
and the $\tau$ lepton (dash-dot-dot-dot),
in units of (c) Im $Z_0$ and (d) Im $\tilde{Z}_0$.  
\label{fig:m01}
}\end{figure}
%

% FIG. 3
\begin{figure}
\caption[]{
The same as in Fig. 2, except that $\tan\beta = 7$.
\label{fig:m07}
}\end{figure}
%

% FIG. 4
\begin{figure}
\caption[]{
The same as in Fig. 4, except that $\tan\beta = 20$.
\label{fig:m020}
}\end{figure}
%

% FIG. 5
\begin{figure}
\caption[]{
The muon EDM generated from one-loop diagrams (dash), 
two-loop diagrams (dash-dot) and their total (solid) 
in units of (a) Im $Z_0$ and (b) Im $\tilde{Z}_0$,  
as a function of $\tan\beta$, with $m_0 =$ 100 GeV.
Also shown are the contributions from two-loop diagrams 
involving the $W$ boson (dot),
the $t$ quark (dash-dot),
the $b$ quark (dash-dot-dot), 
and the $\tau$ lepton (dash-dot-dot-dot),
in units of (c) Im $Z_0$ and (d) Im $\tilde{Z}_0$.  
\label{fig:tanb}
}\end{figure}
%

% FIG. 6
\begin{figure}
\caption[]{
The ratio of ($d_\mu/d_e$) to ($m_\mu/m_e$) 
in units of (a) Im $Z_0$ and (b) Im $\tilde{Z}_0$, 
as a function of $\tan\beta$, 
for $m_0 =$ 100 GeV (solid), 200 GeV (dash) and 300 GeV (dash-dot).
\label{fig:dmde}
}\end{figure}
%

% FIG. 7
\begin{figure}
\caption[]{
The maximal EDM allowed by unitarity (Eq.~\ref{eq:dMAX}) 
for (a) the muon EDM and (b) electron EDM, 
as a function of $\tan\beta$ 
for $m_0 = 100$ (dash), 200 (solid) and 400 GeV (dash-dot). 
Also shown is the experimental upper limit (U.L.) on the electron EDM 
at 95\% C.L. (dot).
\label{fig:dMAX}
}\end{figure}
%

%-----------------------------------------------------------------------
%   END DOCUMENT
%-----------------------------------------------------------------------
\end{document}